\documentclass[twocolumn,prl,aps]{revtex4}
\usepackage{graphics}
\begin{document}
\title{Comment on ``Collapse of Coherent Quasiparticle States in 
$\theta$-(BEDT-TTF)$_2$I$_3$ Observed by Optical Spectroscopy''}
\maketitle

Recently, Takenaka {\it et al.}\cite{TakenakaPRL} reported that
the resistivity $\rho(T)$ of $\theta$-(BEDT-TTF)$_2$I$_3$ ($\theta$-ET) 
exceeds the Ioffe-Regel resistivity, $\rho_{IR}$, by a factor 
of 50 at large temperatures $T$ (``bad metal''). This was 
ascribed to strong correlation. We argue that the optical 
conductivity $\sigma(\omega)$ implies that correlation is not
very strong, and that correlation gives no general strong
suppression of $\sigma(\omega)$. The large $\rho(T)$ is primarily
due to a downturn in $\sigma(\omega)$ at small $\omega$, earlier
emphasized by Takenaka {\it et al.} \cite{TakenakaPRB} as the
explanation for bad metal behavior of high-$T_c$ cuprates. We argue, 
however, that for cuprates strong correlation is the main effect. 
The data of Takenaka {\it et al.}\cite{TakenakaPRL} puts $\theta$-ET 
in a new class of bad metals.           

\begin{figure}
\centerline{
\rotatebox{-90}{\resizebox{!}{3.5in}{\includegraphics{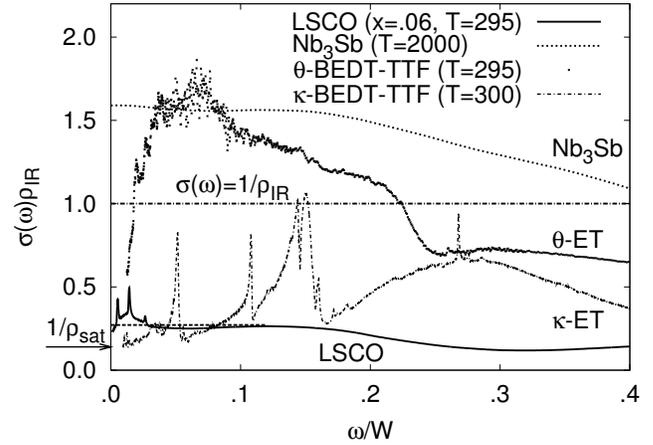}} }}
\caption{\label{fig1}$\sigma(\omega)\rho_{IR}$ as a function of
$\omega/W$ for Nb$_3$Sb (theory) \cite{RMP}, LSCO  
\cite{TakenakaPRB}, $\theta$-ET \cite{TakenakaPRL} and $\kappa$-ET
\cite{Dressel}, where (the noninteracting) $W$ is 8, 3.2, 0.7 and 
1.0 (unrealistically large to show all main features) eV, respectively. 
The interband transitions of LSCO for $\omega/W\gtrsim 0.3$ and sharp 
phonon structures are not of interest here. The dashed curve indicates 
a $\sigma(\omega)$ of LSCO without downturn. The Ioffe-Regel condition implies 
$\sigma(0)\rho_{IR}\gtrsim 1$. All systems are fully in the incoherent limit.}  
\end{figure}

To put $\theta$-ET into context, we discuss a theory of resistivity
saturation \cite{RMP}. We use an  f-sum rule, relating 
$\int \sigma(\omega)d\omega$ 
to the electron hopping energy, $E_K$. We assume that $T$ 
is so large that the Drude peak is smeared out and that $\sigma(\omega)$  
varies smoothly over the band width $W$ (shown for Nb$_3$Sb 
in Fig. \ref{fig1}). We obtain  
\begin{equation}\label{eq:1}
\sigma(0)={\gamma \over W}\int_0^{\infty} \sigma(\omega)d \omega
\sim     {|E_K|\over W},
\end{equation}
where $\gamma \sim 1-2$. Assuming i) noninteracting electrons and  
ii) $T \ll W$, we estimate $E_K$. Inserting $E_K$ in Eq.~(\ref{eq:1}) 
gives $\sigma(0)$ and a quantum-mechanical derivation of the Ioffe-Regel 
condition, $l\gtrsim d$, where $l$ is the apparent mean free path 
and $d$ is a typical atomic separation \cite{RMP}. 

Assumption i) is invalid for cuprates, and  correlation drastically 
reduces $|E_K|$. Estimating $E_K$, we obtained the saturation resistivity 
$ \rho_{\rm sat}$ for La$_{2-x}$Sr$_x$CuO$_4$ (LSCO) 
\cite{RMP}. Experimental data do not exceed $\rho_{\rm sat}$,
and $\rho(T,x)$ curves approaching $\rho_{\rm sat}$ appear to      
saturate.  Fig.~\ref{fig1} shows that strong correlation reduces 
$\sigma(0)$ of  LSCO ($x$=0.06, $T$=295 K) by a factor of four relative 
to $1/\rho_{IR}$, with room for further reduction to $1/\rho_{\rm sat}$ 
with increasing $T$.

Assumption ii) is invalid for A$_3$C$_{60}$ (A=K, Rb). This leads to a     
violation of the Ioffe-Regel condition due to coupling to intramolecular
phonons \cite{RMP}.

For the quarter-filled $\theta$-ET, data for $\int \sigma(\omega)d\omega
\sim E_K$ for small $T$ suggest that correlation does 
not strongly reduce $|E_K|$, in strong contrast to cuprates. As a result, 
$\sigma(\omega)\rho_{IR} \sim 1$, comparable to Nb$_3$Sb, except for 
very small $\omega$. As in the cuprates, there is a downturn in 
$\sigma(\omega)$ at small $\omega$ (``dynamical localization''). While 
this is a small effect
in cuprates, reducing $\sigma(0)$ by about 10-20 $\%$ in Fig.~\ref{fig1} 
(compared with dashed line), the $\theta$-ET data \cite{TakenakaPRL} 
imply a reduction of $\rho(T$=295) by a factor of 50-100. Thus the 
absence of such a downturn is a third assumption (iii) for deriving 
the Ioffe-Regel condition, putting $\theta$-ET in a third class of 
bad metals. The small energy scale of the downturn, $\sim T$,  
raises  questions why the structure is not thermally smeared out.

Calculations \cite{RMP} for A$_3$C$_{60}$, emphasizing coupling to phonons,
show that as $T$ is increased $\int_0^{\omega_{m}}\sigma(\omega)d\omega$
is reduced and reaches its limiting value for larger $\omega_{m}$.
This is similar to $\theta$-ET but very different from cuprates 
(cf. Fig. 4 in Ref. \onlinecite{TakenakaPRL} and inset of Fig. 2 
in Ref. \onlinecite{TakenakaPRB}), suggesting a role for phonons 
in  $\theta$-ET.

Fig. 1 shows that $\sigma(\omega)\rho_{IR}$ is suppressed 
for the ``half-filled'' (BEDT-TTF)$_2$Cu[N(CN)$_2$]Br$_{0.85}$Cl$_{0.15}$
($\kappa$-ET) due to correlation
reducing $|E_K|$ and expanding the energy scale. There is also a
downturn, although less dramatic and on a larger energy scale than 
for $\theta$-ET . Thus both assumptions i) and iii) are moderately violated.

\noindent
O. Gunnarsson and K. Vafayi

\noindent
Max-Planck-Institut f\"ur Festk\"orperforschung,

\noindent
D-70506 Stuttgart, Germany

\

\noindent
Pacs:72.10.-d, 72.80.Le,72.80.Ga

\end{document}